\documentclass[]{pasj02} 
\usepackage[switch,mathlines]{lineno} 
\usepackage[switch,mathlines]{lineno}

\jyear{2024}
\Received{}
\Accepted{}

\newcommand\NSmodsrc{4U~1630$-$472} 

\newcommand\NShigh{GX~340+0}
\newcommand\NHmol{{\rm N(H}_2{\rm )}}
\newcommand\NH{N_{\rm H}}
\newcommand\NHI{N_{\rm HI}}
\newcommand\SI{{\rm S{\sc i}}}

\newcommand\col{~cm$^{-2}$}

\begin{document} 

\title{ XRISM insights for interstellar Sulfur }

\author{
 L\'ia \textsc{Corrales},\altaffilmark{1}\altemailmark\orcid{	
0000-0002-5466-3817} \email{liac@umich.edu} 
 Elisa \textsc{Costantini},\altaffilmark{2,3}\orcid{0000-0001-8470-749X} 
 Sascha \textsc{Zeegers},\altaffilmark{4}\orcid{0000-0002-8163-8852} 
 Liyi \textsc{Gu},\altaffilmark{2}\orcid{0000-0001-9911-7038} 
 Hiromitsu \textsc{Takahashi},\altaffilmark{5}\orcid{0000-0001-6314-5897} 
 David
\textsc{Moutard},\altaffilmark{1}\orcid{0000-0003-1463-8702} 
 Megumi \textsc{Shidatsu},\altaffilmark{6}\orcid{0000-0001-8195-6546} 
 Jon M.~\textsc{Miller},\altaffilmark{1}\orcid{0000-0003-2869-7682} 
 Misaki \textsc{Mizumoto},\altaffilmark{7}\orcid{0000-0003-2161-0361} 
Randall K.
\textsc{Smith},\altaffilmark{8}\orcid{0000-0003-4284-4167} 
Ralf
\textsc{Ballhausen},\altaffilmark{9,10}\orcid{0000-0002-1118-8470} 
Priyanka
\textsc{Chakraborty},\altaffilmark{8}\orcid{0000-0002-4469-2518} 
Mar\'ia 
\textsc{D\'iaz Trigo},\altaffilmark{11}\orcid{0000-0001-7796-4279} 
Renee
\textsc{Ludlam},\altaffilmark{12}\orcid{0000-0002-8961-939X} 
Takao
\textsc{Nakagawa},\altaffilmark{13,14}\orcid{0000-0002-6660-9375} 
Ioanna
\textsc{Psaradaki},\altaffilmark{15}\orcid{0000-0002-1049-3182} 
Shinya
\textsc{Yamada},\altaffilmark{16}\orcid{0000-0003-4808-893X} 
and 
Caroline A. 
\textsc{Kilbourne}\altaffilmark{9}\orcid{0000-0001-9464-4103} 
}
\altaffiltext{1}{University of Michigan, Department of Astronomy, Ann Arbor Michigan, US}
\altaffiltext{2}{Netherlands Institute for Space Research 
(SRON), Leiden, NL}
\altaffiltext{3}{University of Amsterdam, Anton Pannekoek Institute for Astronomy, Amsterdam, NL}

\altaffiltext{4}{European Space Agency, European Space Research and Technology Centre, 
Noordwijk, NL}
\altaffiltext{5}{Hiroshima University, Department of Physical Science, Hiroshima, Chugoku, JP}
\altaffiltext{6}{Ehime University, Department of Physics, Matsuyama, Ehime, JP}
\altaffiltext{7}{Science Research Education Unit, University of Teacher Education Fukuoka, 
Munakata, Fukuoka, JP}
\altaffiltext{8}{Harvard Center for Astrophysics, Cambridge Massachusetts, US}
\altaffiltext{9}{NASA Goddard Space Flight Center, Greenbelt, MD, US}
\altaffiltext{10}{University of Maryland College Park, College Park, MD, US}
\altaffiltext{11}{European Southern Observatory, Garching bei Muenchen, DE}
\altaffiltext{12}{Wayne State University, Department of Physics, Detroit, US}
\altaffiltext{13}{Inst Space \& Astronautical Science, Japan Aerospace Exploration Agency, 
Sagamihara, Kanagawa, JP}
\altaffiltext{14}{Advanced Research Laboratories, Tokyo City University, Setagaya-ku, Tokyo, JP}
\altaffiltext{15}{Massachusetts Institute of Technology, Kavli Institute for Astrophysics and Space Research, Cambridge Massachusetts, US}
\altaffiltext{16}{Rikkyo University, Department of Physics, Toshima-ku, Tokyo, JP}


\maketitle

\begin{abstract}

The X-ray Imaging Spectroscopy Mission (XRISM) provides the best spectral resolution with which to study Sulfur (S) $K$-shell photoabsorption features from the interstellar medium (ISM). For the first time, we demonstrate the high-signal detection of interstellar atomic S{\sc ii} $K$-beta absorption in the spectrum of X-ray binaries (XRBs) \NSmodsrc\ and \NShigh. The persistence of this feature across multiple instruments, targets, and flux states implies that it is interstellar in nature. We measure the S{\sc ii} K$\beta$ line centroid at $2470.8 \pm 1.1$~eV after including systematic uncertainties. We also find that the most recently published high resolution S{\sc ii} absorption template requires a systematic energy scale shift of $+7-8$~eV, which is comparable to the level of disagreement among various atomic modeling procedures. The XRISM 300~ks observation of \NShigh\ provides unprecedented signal-to-noise in the S K region, and we find evidence of residual absorption from solid S in the spectra of \NShigh. Absorption templates from three Fe-S compounds, troilite (FeS), pyrrhotite (Fe$_7$S$_8$) and pyrite (FeS$_2$), provide equally good fits to the residuals. Even though we are not able to distinguish among these three compounds, they provide equal estimates for the abundance of S locked in dust grains. Having accounted for both the gaseous and solid S in the \NShigh\ sightline provides us with a direct measurement of S depletion, which is $40\% \pm 15\%$. Our depletion measurement provides an upper limit to the fraction of interstellar Fe bound in Fe-S compounds of $< 25\%$, which is consistent with prior studies of Fe-S compounds via Fe $L$-shell absorption. Both XRBs in this study are at a distance of approximately 11~kpc and on the opposite side of the Galactic disk, suggesting that this value could represent the average S depletion of the Milky Way when integrated across all phases of the ISM.

\end{abstract}


\section{Introduction}

High resolution X-ray spectroscopy provides a direct means to study heavy elements in the interstellar medium (ISM) in \textit{both} gas and solid form via photo-absorption and scattering of deep (K and L-shell) electron energy levels, which are imprinted on the spectra of bright Galactic X-ray binaries (XRBs) \citep{Costantini2022}. The exact photoabsorption signature of these elements changes with gas ionization state as well as the chemical composition of any solid in which the atom is bound. 
XRISM provides the best spectral resolution with which to study the Sulfur (S) K-shell photoabsorption resonances at 2.45~keV.

The element S is of key interest to the development of life \citep{Huxtable1986, Walsh2020, Olson2021} 
and is also one of the more mysterious elements in interstellar chemistry. While the observed abundance of S in diffuse phases of the ISM appear equal to the expected cosmic abundances \citep{Sofia1994, Savage1996, Martin-Hernandez2002, Howk2006, Neufeld2015}, S in high-column ISM sight-lines can be depleted by a factor of several \citep{Jenkins2009, Psaradaki2024} and can reach up to 99\% depletion in the densest regions of molecular clouds (MCs) \citep{Woods2015, Laas2019, Hily-Blant2022, Fuente2023}. 
In areas where S is depleted, only a small fraction of S is observed in S-bearing molecules and ices \citep{Oppenheimer1974, Tieftrunk1994, Palumbo1997, Vastel2018, Riviere-Marichalar2019, Fuente2019}, implying that the ``missing'' S is in dust grains \citep{Woods2015}.
The specific abundant species of molecular and solid S in a MC depends on factors such as MC age, shock conditions, and the chemical composition of the grains in the cloud \citep{Charnley1997, Hatchell1998, Viti2001, Wakelam2004, Wakelam2011, Laas2019, Perrero2024}. The dominant hypotheses for the reservoirs of solid-phase S include ices formed on the surfaces of refractory grains, S allotropes (e.g., S$_8$),  complex S ``residues'' left behind by the desorption of icy dust grain mantles, and direct adsorption onto silicate dust grains \citep{Caselli1994, Smith1991, vanderTak2003, Jimenez-Escobar2011, Jimenez-Escobar2014, Woods2015, Laas2019, Fuente2019, Shingledecker2020, Cazaux2022, Perrero2024}.

Identifying the chemical composition of S-bearing grains can help us understand the formation of our own Solar System and other planetary systems \citep{Kama2019}. For example,  Solar System comets are suspected to host materials that are representative of the pre-solar nebula, and S has been observed as S-O, S-S, and S-C ices in Comet 67P/Churyumov–Gerasimenko \citep{Calmonte2016}.
The potential uptake of interstellar S by metals such as Fe is also particularly interesting because Fe-S compounds are found in meteorites, interplanetary dust particles, and some cometary dust particles \citep{Bradley2010, Hanner2010, Mann2010}. 
However, all X-ray absorption studies to date evaluating interstellar neutral iron (Fe) L-shell photoabsorption features, which provide a means to identify the solid phase compounds of Fe-bearing dust grains, have not found evidence of Fe-S compounds \citep{Psaradaki2023, Westphal2019, Corrales2024}. 
Other refractory elements, such as Mg, are more efficient at binding with S \citep{Perrero2024}. The molecules MgS and NaS, recently detected in the Galactic Center molecular cloud G+0.693-0.027, could have arisen from dust destruction in cloud-cloud collisions, a potential insight into the mineralogy of S-bearing dust 
\citep{Rey-Montejo2024}.

Atomic \SI\ may only account for some of the S in molecular clouds, depending on the density and age of the MC. Via the mid-infrared, \citet{Fuente2024} detected atomic \SI\  at the edges of the photodissociation region between the ionizing H\,{\sc ii} region of the Orion Nebula and the denser molecular region of the Orion Bar. However, chemical models by \citet{Fuente2024} indicate that S\,{\sc i} is the dominant form of S only in a narrow range of $A_V \sim 3-5$ between an ionizing front and colder molecular material.  

We also consider more highly ionized forms of S, which might exist in hotter regions of the ISM. Taking into consideration the average number densities and filling factor of the various ISM phases as described by \citet{DraineISM}, the column number density of a deep sight-line would typically contain 61\% neutral H\,{\sc i}, 36\% molecular clouds, 2.5\% warm ionized H\,{\sc ii}, and 0.25\% hot ($\gtrsim 10^5$~K) material. The ionization potential of S{\sc ii} is 23.3~eV, so producing S{\sc iii} requires a source of extreme UV irradiation, such as an O star. However, in the majority of the warm ionized medium, S\,{\sc ii} is the more abundant form of S when compared to H\,{\sc ii} regions, indicating that the source of ionization in the diffuse warm ISM is generally softer \citep{Madsen2006}. Thus we expect a negligible contribution of S\,{\sc iii} or higher ionization towards the ISM absorption spectrum. Outside of molecular clouds, S\,{\sc ii} is expected to be the dominant form of S throughout the diffuse ISM  \citep{Savage1996, Laas2019}.

\section{Interstellar Sulfur Absorption Cross-Sections}

S\,{\sc ii} is expected to be the dominant ion of atomic S throughout the diffuse ISM. The ionization energy of S\,{\sc i} is only 10.36~eV, which is lower than that of H\,{\sc i}, so that S in the cold and warm neutral regions of the ISM is not shielded from ionization by the 
interstellar radiation field. Spectroscopic studies of the ISM confirm this; S\,{\sc ii} abundances as measured by UV spectroscopy are consistent with solar abundances in the cool and warm components of the local ISM \citep{Savage1996}. However, many sightlines with column densities $\NH > 3 \times 10^{19}$~cm$^{-2}$ show S\,{\sc ii}  to be depleted by a factor of $3-10$ \citep{Jenkins2009}. Nearly all Galactic XRBs meet this criterion. For example, S is observed to be depleted by a factor of 3 in the diffuse sight line of Cygnus~X-2 \citep{Psaradaki2024}, which has $\NH = 2 \times 10^{21}$~\col\ as estimated from the X-ray band.

S\,{\sc ii} is difficult to study with UV spectroscopy, where its only observable feature is a $1250$\AA\ triplet that is usually saturated \citep{Jenkins2009}. In the X-ray, gas-phase S can be probed through photoabsorption of inner-shell electrons. Interstellar S\,{\sc ii} is expected to show strong K$\beta$ line absorption in addition to a complex of Rydberg series absorption lines leading to the $n=1$ photoelectric ionization energy
around 2.5~keV. Thus X-rays provide another means to study interstellar S and its depletion, and are expected to be 
prominent for ISM column densities $\NH \gtrsim 5 \times 10^{22}$~\col.

\begin{figure}
    \centering
    \includegraphics[width=1.0\linewidth, trim={0 0 0 0}]{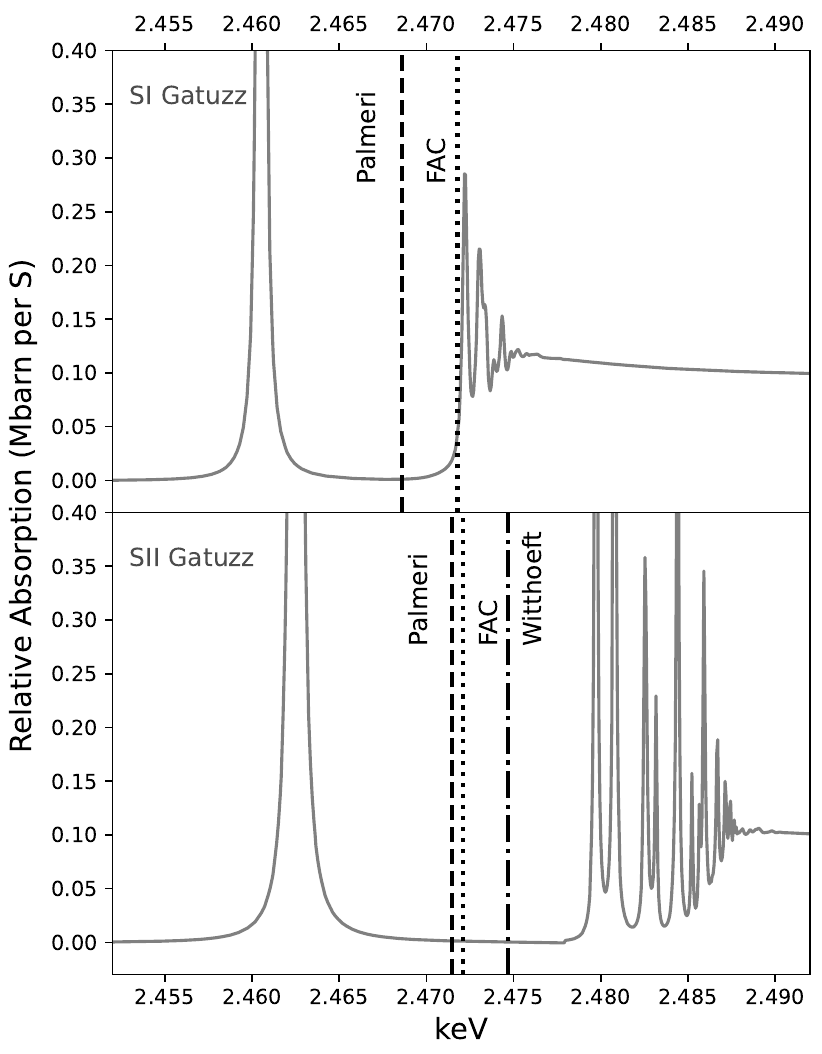}
    \caption{Predicted absorption cross-sections of atomic S{\sc i} (top) and S{\sc ii} (bottom). \citet{Gatuzz2024} provides the most detailed state-of-the-art cross-section templates available (grey). The overlaid vertical lines give predictions for the S K$\beta$ lines of each ion from \citet{Palmeri2008}, the Flexible Atomic Code (FAC) and \citet{Witthoeft2011}.
    In all cases, the predictions for the position of the K$\beta$ features differ by several to 10~eV. \\
    }
    \label{fig:xsects}
\end{figure}

At present there are no laboratory measurements of S{\sc II} K-shell absorption cross sections, transition energies, or ionization energies; we must rely on theory.
Figure~\ref{fig:xsects} shows the most recent, state-of-the-art theory calculation for atomic S\,{\sc i} and S\,{\sc ii} cross-sections using a modified R-matrix method that accounts for Auger broadening \citep{Gorczyca2000, Gatuzz2024}. Overlaid on these plots is the expected position of the K$\beta$ absorption lines as predicted by various works:  the calculations of \citet{Palmeri2008} using the pseudo-relativistic Hartree-Fock (HFR) method, 
the R-matrix calculations of \citet{Witthoeft2011}, 
and the results of running the Flexible Atomic Code (FAC, version 1.1.5). The FAC calculation we performed includes a complete set of inner-shell levels up to $n=5$. Relativistic electron-electron interactions, incorporating both Coulomb and Breit terms, are accounted for within the atomic central potential. Higher-order electronic interactions, which are challenging to describe analytically, are approximated through configuration mixing of the bound states within each $n$ group. A comparison of  S{\sc ii} K$\beta$ line predictions from HFR and FAC are listed in the Appendix (Table~\ref{tab:AtomicData}). See \citet{Gatuzz2024} for a more detailed comparison between their cross-sections and those presented by \citet{Witthoeft2011}. In all cases, the predictions for the position of various K$\beta$ features differ by 5-10~eV.

For the solid phase, we use three absorption cross-sections of three Fe-S compounds: troilite (FeS), pyrrhotite (Fe$_7$S$_8$), and pyrite (FeS$_2$) used in \citet{Costantini2019} and \citet{Gatuzz2024}. 
The optical constants for these compounds were extracted from laboratory absorption measurements\footnote{See the European Synchrotron Radiation Facility S K-edge XANES spectra database, \citet{BonninMosbah2002}, \citet{Costantini2019} and references therein.} and used to compute the absorption cross-section via Mie theory, which accounts for self-shielding \citep{Wilms2000, Corrales2016}. The integrated cross-sections were computed assuming a power-law distribution of dust grain sizes following \citet{Mathis1977}. For all three compounds, the oxidation state of Sulfur is S$^{2-}$, making it Ar-like. This makes it so that the $n=3$ shells are full, which prohibits a strong K$\beta$ absorption resonance. 
Due to the changes in energy potential that may be caused by the oxidation state and the alteration of energy level structure caused by incorporation into solid, we don't necessarily expect the resonant spectral absorption features of solids to align with the $K\gamma$ features of gaseous S. Additionally, the solid cross-sections demonstrate a few eV spread among the first absorption peaks of these three compounds. The cross-sections are shown in Figure~\ref{fig:results}.

\section{Datasets}
\label{sec:data}

\subsection{\NSmodsrc}

The Galactic black hole X-ray binary (BHXRB) \NSmodsrc\ undergoes transient outbursts that begin with a rising hard-state that transitions to a soft-state before decaying to quiescence \citep{Miyamoto1995, Belloni2010, Maccarone2003, Kalemci2013, Dincer2014}. Outbursts from \NSmodsrc\ exhibit different spectral evolution behavior compared to typical BHXRB transients \citep{Abe2005, Tomsick2005, Tomsick2014}. \citet{Kalemci2018} investigated whether spectral anomalies might be attributed to dust obscuration and the corresponding spectral softening from a time-delayed dust echo \citep{Heinz2015, Heinz2016}. 
They used \textit{Chandra} to identify a dust scattering ring, attributable to molecular cloud MC-79, which constrains the distance of \NSmodsrc\ to $11.5\pm 0.3$~kpc.\footnote{Due to degeneracies in interpreting distances to MCs using line-of-sight velocity, there are two possible values for the distance to \NSmodsrc. We follow  \citet{Kalemci2018} in adopting the larger distance due to extensive modeling of the dust ring echo surface brightness profile.} The integrated CO emission from this cloud implies $\NHmol = 8.8(\pm 2.6) \times 10^{21}$\col, but various other molecular clouds in the sightline bring the total to $\NHmol = 20.0(\pm 6.1) \times 10^{21}$\col. Combining the molecular gas column with the 21-cm integrated abundance $\NHI \approx 2 \times 10^{22}$\col\ from \citet{Kalberla2005}
implies that this sightline has $\NH \geq 6.0(\pm 1.2) \times 10^{22}$\col. 
Since the 0.3-10~keV X-ray band is more sensitive to absorption from C, O, and heavier elements, X-ray extinction is better parameterized by the total column density of H-nuclei in the sightline ($\NH$). Spectral fitting of \NSmodsrc\ throughout the literature has suggested $N_H \approx 5-15 \times 10^{22}$\col \citep{Tomsick2014}, $8.3 - 10 \times 10^{22}$\col \citep{Neilsen2014}, and $9.7 \times 10^{22}$\col \citep{Zeegers2019}. 
Likely the most accurate measurement of the H column comes from \citet{DiazTrigo2013}, who tested the effect of various X-ray continuum model assumptions and found they were all consistent with $N_H = 7-9 \times 10^{22}$\col. The large range of measured $N_H$ values has led to the suggestion that some level of the absorption is variable and thereby associated with the XRB environment itself \citep{DiazTrigo2013, Neilsen2014}.

The XRISM observation (OBSID: 900001010) was performed on 16--17 February 2024 as a target of opportunity to catch the decaying phase of the outburst. XRISM observed the source at relatively dim phase with a 2--10 keV intensity of $\sim 50$ mCrab, before the transition from the high/soft state, where the source spectrum is dominated by the thermal disk emission, to the low/hard state, where a hard power-law shaped spectrum is seen. The microcalorimeter \textit{Resolve} was operated in the non-filter mode under the closed gate valve condition and its net exposure was $\sim$ 50 ks. The nominal data screening was performed, and the ``Hp'' spectra were extracted, which has the best possible energy resolution. Some pixels were removed to avoid the contamination of the ``pseudo-Ls'' events. The redistribution matrix file (RMF) was calculated with the ``X'' option, which includes all the known instrumental effects. See Miller et al. (ApJL in review) for the full details regarding the data reduction and scientific background.

We also examined archival \textit{Chandra} HETG spectra of \NSmodsrc, 
OBSIDs 13714-13717, which are a set of four 30~ks exposures obtained during the 2012 outburst.  The observations were obtained at similar flux levels, permitting their addition.  The data were downloaded from the public Chandra archive, and reprocessed using the CIAO script \texttt{chandra\_repro}.  First-order HEG spectra were then created by running the tools \texttt{tgdetect}, \texttt{tg\_create\_mask}, \texttt{tg\_resolve\_events}, and \texttt{tgextract}.  The default parameter values were used in each tool with one exception: the ``width\_factor\_hetg'' parameter within \texttt{tg\_create\_mask} was changed from a default value of 35 to a value of 18 in order to better separate orders at low dispersion angle. 
Response files were then created for each grating arm using the tools \texttt{mkgrmf} and \texttt{fullgarf}.  Finally, the spectra and responses were combined using the CIAO script \texttt{combine\_grating\_spectra}.  After filtering each observation, the combined exposure is 116.9~ks.  

\begin{longtable}{l c c c c c}
    \caption{Parametric fit to the S~K absorption features}
    \label{tab:parametric}
    \hline\noalign{\vskip3pt} 
        Observation & $K\beta$ position & N$_{SII}$ & EW & $K$-edge position & $K$-edge strength \\   [2pt] 
            & (eV) & ($10^{16}$~cm$^{-2})$ & (eV) & (eV) & (Max $\tau$) \\  [2pt] 
    \hline\noalign{\vskip3pt} 
    \endhead
    \hline\noalign{\vskip3pt} 
    \endfoot
        \NSmodsrc\ Combined HEG & $2471.0 \pm 0.4$ & $153_{-42}^{+47}$ & $1.79_{-0.34}^{+0.42}$ & $2483.9_{-1.6}^{+1.0}$ & $0.184 \pm 0.003$ \\
        \NSmodsrc\ High Flux (XRISM) & $2470.7 \pm 1.4$ & $95_{-42}^{+55}$ & $1.37_{-0.54}^{+0.75}$ & $2480.9_{-2.6}^{+3.1}$ & $0.184 \pm 0.008$ \\
        \NSmodsrc\ Low Flux (XRISM) & $2470.6_{-1.1}^{+1.2}$ & $64_{-50}^{+77}$ & $1.05_{-0.78}^{+0.12}$ & $2480.9_{-2.6}^{+3.1}$ & $0.184 \pm 0.008$ \\
    \hline\noalign{\vskip3pt} 
        GX~340+0 \textit{Chandra} 18085 HEG-1 & $2470.3 \pm 0.7$ & $116_{-46}^{+65}$ & $1.55_{-0.58}^{+0.84}$ & $2484.8 \pm 2.9$ & $0.140 \pm 0.006$ \\
        GX~340+0 \textit{Chandra} 19450 HEG-1 & $2470.4 \pm 0.4$ & $119_{-34}^{+42}$ & $1.57_{-0.40}^{+0.50}$ & $2477.4_{-2.0}^{+1.6}$ & $0.199 \pm 0.004$ \\
        GX~340+0 \textit{Chandra} 20099 HEG-1 & $2470.6_{-0.5}^{+0.7}$ & $125_{-35}^{+42}$ & $1.75_{-0.58}^{+0.87}$ & $2482.9_{-1.5}^{+1.2}$ & $0.170 \pm 0.004$ \\
        GX 340+0 XRISM 300002010 (Obs 1) & $2470.5 \pm 0.3$ & $117_{-24}^{+26}$ & $1.55_{-0.23}^{+0.27}$ & $2482.0_{-1.2}^{+0.8}$ & $0.140 \pm 0.002$ \\
        GX 340+0 XRISM 300002020 (Obs 2) & $2471.0 \pm 0.3$ & $79_{-17}^{+19}$ & $1.27_{-0.21}^{+0.23}$ & $2480.8_{-1.4}^{+1.8}$ & $0.133 \pm 0.002$ \\
        GX 340+0 XRISM combined & $2470.8 \pm 0.2$ & $95_{-17}^{+18}$ & $1.41_{-0.16}^{+0.17}$ & $2481.7_{-1.0}^{+0.9}$ & $0.134 \pm 0.001$ \\
\end{longtable}

\subsection{\NShigh}

\NShigh\ is a neutron star low-mass X-ray binary (LMXB) that presents the best opportunity to study interstellar S, due to its high luminosity $>0.5 L_{\rm Edd}$ \citep{seifina13}, smooth spectrum with relatively few emission and absorption features, and high obscuring ISM column. The estimated distance to \NShigh\ is $11\pm3$ kpc \citep{miller93, Gilfanov2003}. The estimated neutral atomic H column, $\NHI = 2.1 \times 10^{22}$ \citep{Kalberla2005}, is similar to that of \NSmodsrc, and by extension one might expect a similar amount of molecular material in the sight line. 
The H column measured from X-ray spectra of \NShigh\ varies from $5.5-6.8 \times 10^{22}$\col \citep{Zeegers2019} to $10.9 \pm 0.1 \times 10^{22}$\col \citep{Cackett2010}. 
The 3D dust maps of \citet{Edenhofer2024} give an $A_V = 1.9$, implying $N_H = 3.8 \times 10^{21}$\col for the ISM within 1.25~kpc of this sight line, which accounts for $<10\%$ of the expected total. 

The brightness and level of ISM obscuration make \NShigh\ the highest signal-to-noise target among all Galactic XRBs with which to study the K-shell spectral features of silicon \citep{Zeegers2019}, sulfur (this work), iron \citep{Rogantini2018}, and other trace refractory elements \citep{Costantini2019}. This target varies in spectral and timing behavior as it moves along the "Z" track of a hardness-intensity diagram \citep{chattopadhyay24}, which is the subject of another work (Ludlam et al, PASJ in prep). We found no evidence for change in the S K-shell absorption features to within $1\sigma$ when we divided the spectra according to the position on the Z track. We thereby opted to use the integrated time-averaged spectra from archival \textit{Chandra} and new XRISM observations for this study.

\NShigh\ was observed twice during the XRISM Performance Verification phase, for 152~ks on 17--21 August 2024 (OBSID: 300002010) and for 153~ks on 31 August -- 04 September 2024 (OBSID: 300002020). The \textit{Resolve} spectrum was extracted following standard procedures.\footnote{https://heasarc.gsfc.nasa.gov/docs/xrism/analysis/abc\_guide/xrism\_abc.html}  
We then extracted the primary high resolution (Hp) events, used the \texttt{rslmkrmf} tool to make the small version of the \textit{Resolve} response matrix, and used the \texttt{xaarfgen} tool to make the area response file for a point source. We found that the large and extra-large response files were more effective for achieving good continuum fits for the soft end of the spectrum, at $E < 2.3$~keV. As described below, for the purposes of this study, the only portion of the continuum fitting that affects our results is the $2.5 - 4$~keV continuum absorption from S. 
During OBSID 300002010, pixel 30 experienced a gain jump that affected the energy scale by about 1.5 eV at 2.5 keV, for approximately 70 ks after the end of the second ADR cycle. The expected impact is a broadening of the line-spread function at 2.5 keV by 0.1 eV over that time interval. However, we found that removing pixel 30 from OBSID 300002010 did not change our best fit parameters for the S~K$\beta$ line, so we continued with this pixel included in the dataset.

We note also that GX 340+0 is extremely bright, and during certain periods of the observation, event loss occurred because the onboard CPU in \textit{Resolve} could not process some events \citep{Mizumoto2025}. This slightly alters the effective exposure time, which changes the absolute flux but does not affect the spectral shape. 
The brightness level of this target significantly also reduces the number of Hp events available from inner four pixels of the detector, making it so that the outer pixels dominate the spectrum. 
We compared the Hp spectrum extracted from the four inner pixels with the Hp spectrum from the remaining 31 pixels and found that the shape of the S~$K$-shell photoabsorption features did not differ between the two. 
Finally, it has been found that bright sources could also affect the gain calibration \citep{Mizumoto2025}. We examined this possibility by fitting for the line centroid of the instrumental Si K$\alpha$ fluorescence feature, finding that it was consistent with measurements of the Si K$\alpha$ feature from XRISM targets with lower count rates (Eckart JATIS 2025, in prep). We conclude that the high count rate of \NShigh\ minimally impacts our study of the S K-shell photabsorption features.

This analysis also includes archival \textit{Chandra} HETG spectra of \NShigh. We  selected OBSIDs 18085, 19450, and 20099 where a sub-windowed observation mode was chosen to mitigate pileup. The reduced spectra were downloaded from \textit{tgcat}\footnote{https://tgcat.mit.edu}. For details of the catalog data reduction pipeline, see \citet{Huenemoerder2011}. All three observations employed a pile-up mitigation strategy that made it so that only the $m=-1$ order HEG (HEG-1) spectrum was available for studying S~$K$-shell absorption features.

\section{Fitting Methods and and Results}

Each spectrum was fit individually using XSPEC v12.14.1 \citep{Arnaud1996} models accessed through the Interactive Spectral Interpretation System \citep{Houck2013}, we binned the data to 100 counts per bin for the continuum fitting process. We modeled the $2.3-10$~keV continuum with either a powerlaw or a black-body disk model \texttt{diskbb} under the effects of ISM absorption with \texttt{tbvarabs} \citep{Wilms2000} and scattering with \texttt{xscat} \citep{Smith2016}. In the case of \NSmodsrc, we set the position of the dust scattering material to align with the MC-79 cloud. For \NShigh, we put the dust at the half-way point along the line of sight. However, across all targets, we found that the choice of \texttt{xscat} model position did not affect our ability to find a good fit statistic for the continuum.

The best-fit $\NH$ column varies by a factor of two depending on the abundance table chosen for computing the \texttt{tbvarabs} model, as well as the position of the dust along the line-of-sight chosen for the \texttt{xscat} model. 
For this work, we opt to ignore the details of the continuum parameters to focus instead on the distinct spectral features of S K. We then left the abundance of S in the \texttt{tbvarabs} model as a free parameter to manage the contribution of S K absorption to the continuum. In all cases, strong line absorption was apparent at energies just below the predicted S K shell absorption edge \citep{Verner1995}, which would be consistent with the SII K$\beta$ line feature.

\subsection{Parametric fit for K shell absorption}
\label{sec:parametric}

After obtaining our initial continuum fit, we removed the binning on the data and reduced the fitting range to $2.3-4$~keV, which was broad enough to capture the S~K absorption continuum. Only the low flux state of the \NSmodsrc\ XRISM observation required binning, for which we set a minimum of 25 counts per bin. We also switched to using Cash statistics and refit the continuum model for the narrower band. Next, we set the S abundance in the \texttt{tbvarabs} model to zero and replaced it with a parametric model, consisting of a Voigt line absorption profile (\texttt{voigtabs}) and the \texttt{edge} XSPEC models. Based on visual inspection of the datasets, we placed limits on the line centroid between 2.460 and 2.480~keV and the position of the edge between 2.475 and 2.495~keV. 

Table~\ref{tab:parametric} shows the results for our parametric fit with $1\sigma$ confidence intervals, and Figure~\ref{fig:spectra} (top) shows the fit spectrum. 
We note that the variations in the best fit \texttt{MaxTau} from the XSPEC \texttt{edge} model may result from intrinsic absorption in the binary, sensitivity to the  S K-shell near-edge structure, or continuum-driven fitting differences. Among the XRISM observations, we observed 2.5\% changes in the best fit continuum parameters, implying random variations of about $0.0035$ in \texttt{MaxTau}, which would make all values in Table~\ref{tab:parametric} consistent with each other. The \textit{Chandra} datasets for \NShigh, though consistent with the XRISM spectra, are lower quality and thus have larger fit uncertainties and larger variations in both the K-edge position and its strength. Thus we are unable draw any meaningful conclusions from the variations in the \texttt{edge} model fits.

We find the best fit \texttt{voigtabs} parameters in Table~\ref{tab:parametric} are more physically meaningful. In particular, the XRISM observations of \NShigh\ provide the strongest constraint on the line position, and the $1\sigma$ errors are on the order of the gain calibration uncertainty for the \textit{Resolve} instrument (0.3~eV, but potentially as large as 1~eV for energies below 5.4~keV, Eckart JATIS 2025, in prep). The absorption line feature appears at the same position in both sources for all datasets, providing strong evidence that it is interstellar in nature. 

The strength of the K$\beta$ absorption line can be used to measure the abundance of S{\sc ii} along the sight line. For the Voigt profile, we used the average value of the Einstein $A$ coefficients as calculated by \citet{Palmeri2008} and FAC (Appendix Table~\ref{tab:AtomicData}), and we set the velocity dispersion to 74~km/s, consistent with that observed from H{\sc i} in the Milky Way \citep{Kalberla2009}. The normalization of the Voigt profile, which is directly proportional to the column density of the absorber, is degenerate with these parameters. To account for uncertainties in these values, we propagated an extra 10\% uncertainty from the $A$ coefficients and 10\% uncertainty from the velocity dispersion\footnote{Increasing the velocity dispersion to 100~km/s led to a 10\% change in the strength of the Voigt absorption profile.} into the final error on the column density measurement. The implied S{\sc ii} column densities from each spectra are reported in Table~\ref{tab:parametric}. Even though the column density of S{\sc ii} was calculated directly from the Voigt profile fit, we also report the measured equivalent width in Table~\ref{tab:parametric} as a point of reference. An order of magnitude estimate for the total optical depth of the SK~$\beta$ line is $\tau \sim 100$, indicating that these measurements are in the flat portion of the curve-of-growth.


\subsection{Fit with S{\sc ii} absorption template}
\label{sec:SIItemplate}

\begin{figure}
    \includegraphics[width=\linewidth, trim={0.5cm 0 0 0}]{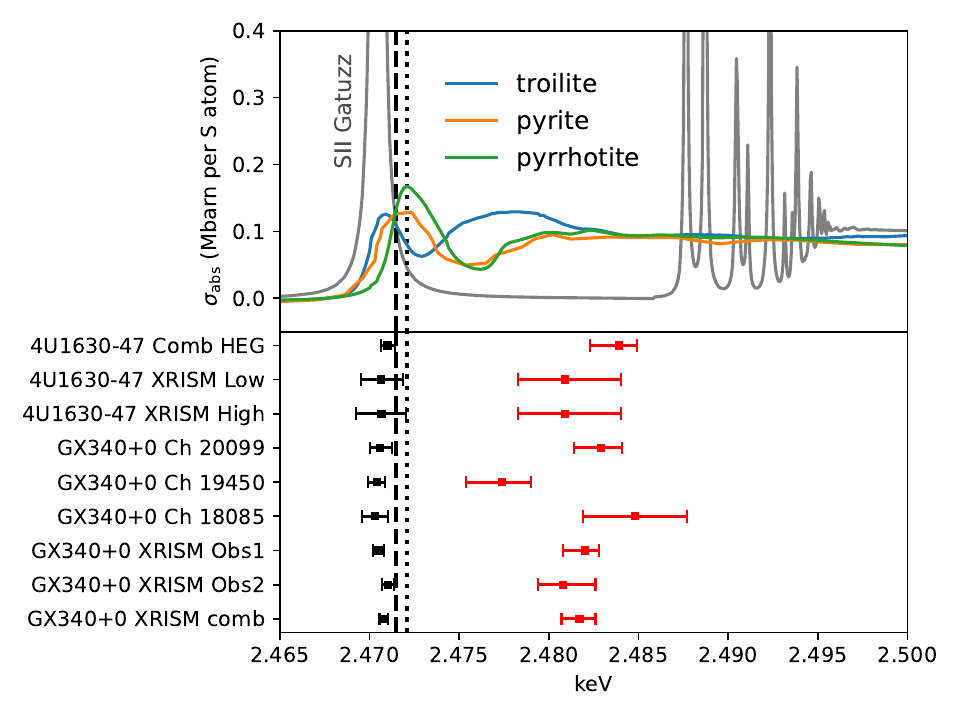}
    \caption{Top: The \citet{Gatuzz2024} S{\sc ii} cross-section with a 7.92~eV blueshift, matching the best fit energy shift for the combined XRISM data. 
    The absorption cross-section templates for three Fe-S compounds used in this work, troilite (FeS, blue), pyrite (FeS$_2$, orange) and pyrrhotite (Fe$_7$S$_8$, green), are overlaid for reference. 
    Bottom: The measured position of the K$\beta$ feature (black) and XSPEC \texttt{edge} model position (red) for each spectrum examined. The dashed and dotted vertical lines show the predicted S{\sc ii} K$\beta$ position from HFR \citep{Palmeri2008} and FAC, respectively. \\
    }
    \label{fig:results}
\end{figure}

\begin{figure}
    \centering
    \includegraphics[width=1.0\linewidth]{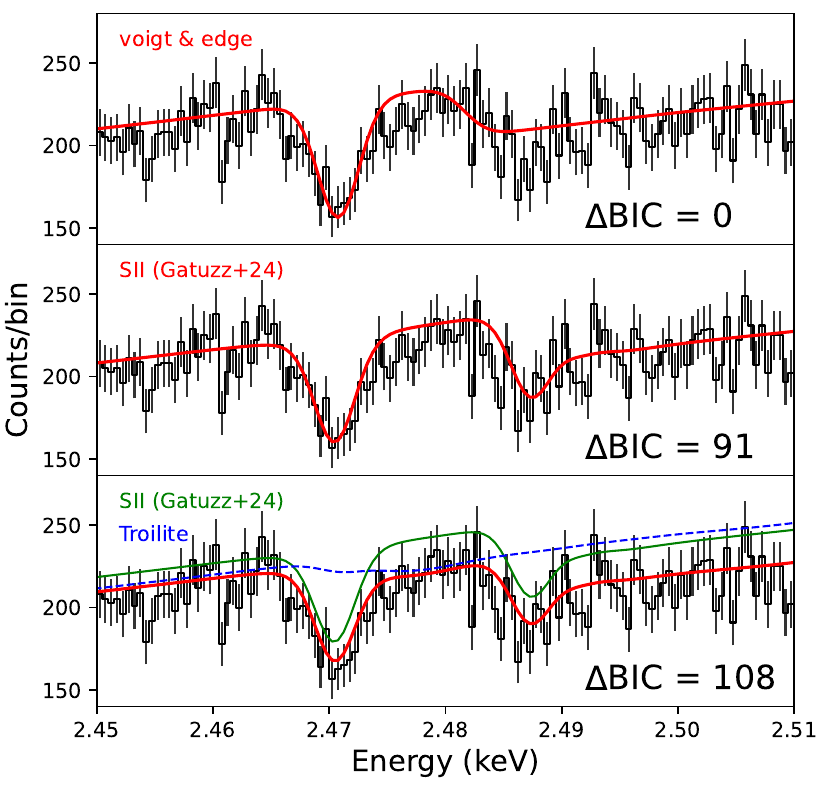}
    \caption{Three simultaneous fits to the two XRISM observations for S K-shell photoelectric absorption in the sight line of GX~340+0 with increasing model complexity. The spectra are evaluated separately but plotted as a combined spectrum for visual clarity.  (Top:) A parametric description of S{\sc ii} absorption using a Voigt profile for the K$\beta$ absorption line and the XSPEC \texttt{edge} model for the photoelectric absorption continuum. (Middle:) A fit with the S{\sc ii} absorption cross-section presented in \citet{Gatuzz2024}, when allowing for a shift in the energy scale. The proposed blueshift of 7.9~eV comes from this dataset. (Bottom:) A fit with the S{\sc ii} template (green solid curve) and absorption from troilite (FeS) dust (blue dashed curve). 
    The change in Bayes Information Criterion relative to the parametric fit (top) is given in each panel, where a positive $\Delta {\rm BIC}$ indicates a more preferred model. \\
    }
    \label{fig:spectra}
\end{figure}

Next we replaced the parametric \texttt{edge} and \texttt{voigtabs} S~K absorption model with the S{\sc ii} absorption cross-section from \citet{Gatuzz2024}. For each target, we initiated the model with a blue-shift of 8~eV to align the S{\sc ii} template K$\beta$ feature by eye with the observed line feature. We then fit for the energy shift and S{\sc ii} abundance, shown with $1\sigma$ confidence limits in Table~\ref{tab:SII}. Figure~\ref{fig:results} shows the measured line and edge positions from Section~\ref{sec:parametric} with respect to a shifted S{\sc ii} template and the K$\beta$ line positions predicted by other theoretical means. The shifted S{\sc ii} absorption template provides a good fit to much of the structure visible in the S K shell photoabsorption region (Figure~\ref{fig:spectra}, middle). The exquisite resolution of XRISM and high signal from this target reveals residual structures arising from solid S-bearing compounds.

\begin{table}[]
    \tbl{Template fit with S{\sc ii} absorption}{
    \begin{tabular}{l c c}
        \hline\noalign{\vskip3pt}
        Observation & $\Delta E$ & N$_{SII}$ \\
            & (eV) & ($10^{16}$~cm$^{-2}$) \\
        \hline\noalign{\vskip3pt}
        \NSmodsrc\ combined HEG & $7.97_{-0.16}^{+0.06}$ & $184_{-13}^{+25} $ \\
        \NSmodsrc\ XRISM high & $7.00_{-0.25}^{+1.07}$ & $171_{-18}^{+26}$ \\
        \NSmodsrc\ XRISM low & $7.92_{-1.53}^{+0.88}$ & $119_{-14}^{+26}$ \\
        \hline\noalign{\vskip3pt}
        GX~340+0 Ch 18085 HEG-1 & $7.99_{-0.15}^{+0.06}$ & $115_{-15}^{+20}$ \\
        GX~340+0 Ch 19450 HEG-1 & $7.87_{-0.07}^{+0.10}$ & $140_{-20}^{^+9}$ \\
        GX~340+0 Ch 20099 HEG-1 & $8.00_{-0.04}^{+0.05}$ & $132_{-9}^{+11}$ \\
        GX~340+0 XRISM Obs 1 & $6.96_{-0.06}^{+0.07}$ & $136_{-14}^{+7}$ \\
        GX~340+0 XRISM Obs 2 & $7.92_{-0.05}^{+0.07}$ & $134_{-14}^{+6}$ \\
        GX~340+0 XRISM combined & $7.92_{-0.05}^{+0.06}$ & $133_{-4}^{+11}$ \\
        \hline\noalign{\vskip3pt}
    \end{tabular}}
    \label{tab:SII}
\end{table}

\subsection{Fit with gas and dust}

The S{\sc ii} template fit includes continuum absorption, which is sensitive to the total abundance of low-ion states of S including those in solids. Additional and unmodeled photoelectric absorption by dust might thereby explain why the total column of S{\sc ii} measured by the S{\sc ii} template (Table~\ref{tab:SII}) is larger than the column measured from the S{\sc ii}~K$\beta$ feature alone (Table~\ref{tab:parametric}). To evaluate this hypothesis, we fit the XRISM \NShigh\ spectra with a model including both the S{\sc ii} template and S-bearing dust templates. Additionally, we aim to investigate if the residuals in the \textit{Resolve} spectrum of \NShigh\ can be used to select among three Fe-S compounds used in \citet{Costantini2019} and \citet{Gatuzz2024}.

Each cross-section for the Fe-S dust templates (Figure~\ref{fig:results}) was computed from the complex index of refraction with the Bohren \& Hoffman Mie scattering algorithm and assuming a powerlaw distribution of dust grains with a slope of $-3.5$ and grain sizes from $0.005-0.25~\mu{\rm m}$. This technique accounts for self-shielding of the dust grain \citep{Wilms2000} and scattering. Self-shielding occurs in grains $\gtrsim 0.3~\mu{\rm m}$ when X-ray light is unable to fully penetrate the grain, so that the observed absorption strength does not scale linearly with the total abundance of the absorbing element. Scattering by dust grains contributes to continuum extinction at a similar order of magnitude to the absorption continuum, but also depends on the imaging resolution of the telescope \citep{Corrales2016}. The scattering effect on the continuum is already accounted for by the \texttt{xscat} model. Scattering also contributes additional spectral features, but we have found that these features only modify the fine structure of the cross-section on the $\leq 5\%$ level. We therefore find the absorption-only cross-section for S-bearing dust grains suitable for measuring abundances. The cross-sections are normalized to the amount of absorption per S atom in the line of sight column.

Table~\ref{tab:dustfit} shows the results from a simultaneous fit using both XRISM exposures. We performed three fits under the assumption solid-phase S is dominated by one of the three proposed compounds. The change in Cash statistic ($\Delta C$) is reported relative to a fit with S{\sc ii} absorption template only. We can see that, while including S-bearing dust improves the fit, there is no one compound that performs better than the others with high statistical significance. Once we  incorporate dust, the S{\sc ii} abundance agrees with that inferred from the Voigt absorption profile (Table~\ref{tab:parametric}). The combined gas and dust abundance of S is also consistent with that measured with the S{\sc ii} template alone (Table~\ref{tab:SII}), which is driven by the continuum absorption. 
Having obtained a full census of the S gas and dust, we now have a direct depletion measurement for this sight line: $40\% \pm 15\%$ of the interstellar S is bound in dust.

We also note that the peak of the S-bearing dust templates overlap with the S{\sc ii} K$\beta$ line, causing the best-fit S{\sc ii} template to shift 0.5~eV to the left in some of the fits. This shift may be real and necessary, but it is also on the order of the gain uncertainty for the XRISM energy scale. Additionally, the absolute energy scale calibration of laboratory measurements for X-ray absorption cross-sections from O K-shell and Fe L-shell have been subject to re-evaluation based on their mismatch with interstellar absorption \citep{Juett2004, Gatuzz2013, Corrales2024}. The solid S K-shell absorption cross-sections in this work were not taken for the purposes of comparison to astrophysical data. Laboratory measurements of S K-shell absorption for the ISM absorption study are forthcoming, and may yield different results for the position of S~K-shell absorption structure in solids.

\begin{longtable}{l c c c c c c}
  \caption{Gas and dust absorption in GX~340+0 (Obs 1 \& 2 simultaneous fit) }\label{tab:dustfit}  
\hline\noalign{\vskip3pt} 
  Compound & NS-solid & N$_{SII}$ & $\Delta E$ & Total N$_S$ & $\Delta C$ 
  & Depletion \\   [2pt] 
  & ($10^{16}$~cm$^{-2}$) & ($10^{16}$~cm$^{-2}$) & (eV) & 
 ($10^{16}$~cm$^{-2}$) &  & (\%) \\  [2pt] 
\hline\noalign{\vskip3pt} 
\endhead
\hline\noalign{\vskip3pt} 
\endfoot
   Pyrrhotite (Fe$_7$S$_8$) & $56_{-9}^{+14}$ & $77_{-9}^{+12}$ & $7.48_{-0.04}^{+0.05}$ & $133_{-13}^{+18}$ & 35 & $42_{-8}^{+12}$ \\
    Troilite (FeS) & $56_{-13}^{+10}$ & $83_{-11}^{+12}$ & $7.44 \pm 0.04$ & $139_{-14}^{+16}$ & 32 & $40_{-10}^{+8}$ \\
    Pyrite (FeS$_2$) & $50_{-7}^{+1}$ & $85_{-12}^{+9}$ & $7.93_{-0.05}^{+0.04}$ & $135_{-14}^{+9}$ & 23 & $37_{-6}^{+3}$ \\ 
\end{longtable}

\subsection{Arguments for the interstellar nature of the absorption}

If the observed strong-line absorption around the S $K$-shell photoabsorption region was intrinsic to the XRBs themselves 
then we would expect to see variability following changes in luminosity or hardness, which has been observed in both sources from this study. 
Our measurements in Figure~\ref{fig:results} show that the suspected S{\sc ii} K$\beta$ line feature appears consistently in the same position across the two different XRB sources, in two different instruments, and across the various accretion states. We examined the spectra of \NShigh\ separated by position on the Z track (Ludlam et al., PASJ in prep) and found that the line feature was present during all states. 
The persistence and strength of the S{\sc ii} $K\beta$ feature is consistent with what we expect for the ISM.

Absorption from photionized absorbers in XRBs can contaminate the search for ISM absorption features in X-ray spectra. The potential contaminants for the S{\sc ii} K$\beta$ feature are strong absorption from He-like S{\sc xv} at 2.461~keV (1s$^2$ $^1$S$_0$ $\rightarrow$ 1s2p $^1$P$_{1}$) and 2.430~keV (1s$^2$ $^1$S$_0$ $\rightarrow$ 1s2s $^3$S$_{1}$) as well as H-like Si{\sc xiv} at 2.506~keV (1s$^1$ $^2$S$_{1/2}$ $\rightarrow$ 4p$^1$ $^2$P ).\footnote{Line positions reported from AtomDB v3.0.9 \citep{Foster2018}.}
To appear at 2.471~keV, the S{\sc xv} features require a velocity shift of $-1000$ or $-5000$~km/s, and the Si{\sc xiv} features require $+4000$~km/s. 
S{\sc xv} and Si{\sc xiv} features are expected to appear strongest with an ionization parameter of $\log \xi \sim 2$. Most photoionized absorption systems found in X-ray binaries have $\log \xi \gtrsim 3$, and the column density must be high ($\NH \sim 10^{23}$~\col) to produce visible absorption from S{\sc xv} and Si{\sc xiv}.

\NSmodsrc\ exhibits variable absorption signatures from photoionized winds \citep{Kuulkers1998, King2014, Trueba2019, Gatuzz2019}. A recent XRISM observation captures variations in photoabsorption from multiple wind components with velocities ranging from a few hundred to $-8000$~km/s, with ionization parameters $\log \xi \gtrsim 3$, and with column densities of $0.3 - 10 \times 10^{22}$~\col (Miller et al., ApJL in review).
The only absorption system of concern has $\log \xi \approx 3$ with a velocity of $-1000$~km/s, has a relatively low column of $\log N_H = 21.4$, and is seen only during the low-flux stage of the XRISM observation. Our study finds evidence of the S{\sc ii} feature during both portions of the \NSmodsrc\ observation, suggesting that the impact of the disk wind absorber is minimal for our study.

A search for photoionized components intrinsic to the \NShigh\ is the subject of another work (Chakraborty et al, in prep). 
To assess any potential degree of contamination in \NShigh, we used a SPEX \texttt{bb * dbb * pion} model \citep{Kaastra1996, Mehdipour2016} and a $\log \xi = 3$ absorber with a velocity shift of $-3000$~km/s, similar to absorption components reported in \citet{Miller2016}. We found that a very high column ($\NH > 10^{23}$~\col) is required for the photoionized plasma to produce any strong absorption features, and only S{\sc XV} provides a significant contribution. The structure of those features does not match that seen in Figure~\ref{fig:spectra}.

As a sanity check, we can compare the total S abundance measured in our sight lines to expectations from measurements of $N_H$ in the literature. 
Both targets have similar estimated H{\sc i} columns of $N_{\rm HI} \approx 2 \times 10^{22}$\col \citep{Kalberla2005} and at least \NSmodsrc\ has an estimated H$_2$ column of $2 \times 10^{22}$\col inferred via CO(1-0) \citep{Kalemci2018}. However, radio measurements of neutral H{\sc i} and the molecular ISM through CO(1-0) emission in these low-latitude regions of the Galaxy suffer from saturation \citep{Dickey1990, Kalberla2009}, so that the abundance determinations technically yield a lower limit for the total H column. 
ISM extinction in the 0.3-10~keV X-ray band is most sensitive to the total abundance of metals like C and O, providing another means to measure total H abundance regardless of its form. Unfortunately, the total H column as measured by X-ray spectra varies significantly depending on the abundance table assumed, treatment of dust extinction, telescope imaging resolution, and the underlying continuum model chosen \citep{Wilms2000, Smith2016, Corrales2016}. This might explain why the measured H column for both binaries has varied significantly from $5 - 15 \times 10^{22}$\col\ throughout the literature (Section~\ref{sec:data}). 
Working backwards, we can assess the H column implied by our measurement for the total S column. Typical S abundances for the local ISM are $12-15 \times 10^{16}$~cm$^{-2}\ ({\rm N}_{\rm H}/10^{22}~{\rm cm}^{-2})$ \citep{Wilms2000, Lodders2003, Lodders2009, Asplund2021}. Given a total S column of $(136 \pm 25) \times 10^{16}$~cm$^{-2}$ for the sight line of \NShigh, the implied H column is $\NH \approx (10 \pm 2) \times 10^{22}$~cm$^{-2}$. 
For \NSmodsrc, the S column measurements from Table~\ref{tab:SII} imply $\NH \approx (11.7 \pm 3.5) \times 10^{22}$\col. If there is a cold neutral absorption system associated with the XRB itself, then it may not be suitable to assume typical ISM abundances for S. Photoelectric absorption from ions in the disk wind might also lead to an overestimate of the S abundance, because the fit is dominated by the continuum data points. 
Nonetheless for both targets, the H column estimated from S abundance are consistent with the literature, mainly due to the large error bars.

\section{Conclusions}
\label{ref:conclusions}

We demonstrate that the exquisite resolution and sensitivity of the XRISM \textit{Resolve} instrument enables the examination of interstellar S in unprecedented detail. This work also demonstrates the first direct measurement of S depletion in the ISM, where both gas and dust are able to be measured directly, which is a unique capability of X-ray spectroscopy.

We have confirmed the existence and position of interstellar S{\sc ii} K$\beta$ line absorption across multiple instruments, targets, and XRB flux states. XRISM, which provides the best spectral resolution at this energy, observes the line at $2470.8 \pm 0.2$~eV. The systematic uncertainty in the gain calibration for energies below 5.4~keV is 1~eV (Eckart et al. 2025).
The location of solid-phase S near-edge structures, if blended with the K$\beta$ feature (as in Figure~\ref{fig:results}), are also a source of systematic uncertainty for the line centroid. Comparing the best fit energy shifts for the S{\sc ii} template in Table~\ref{tab:SII} and Table~\ref{tab:dustfit} demonstrate a difference of about 0.5~eV. This leads us to estimate a 1.1~eV systematic uncertainty overall. These potential variations are also within the uncertainties on the velocity dispersion of the ISM in the galactic plane. 

In an initial study of X-ray absorption by S in the ISM, \citet{Gatuzz2024} did not consider systematic shifts to the atomic cross-sections. As a result, their measurement for S absorption in the \NShigh\ sight line consisted mostly of S{\sc i} and S{\sc iii} with a total abundance $N_S = (102 \pm 11) \times 10^{16}$~cm$^{-2}$ and provided an upper limit on the S{\sc ii} abundance as $< 4\%$ of the total sight line. However, based on UV observations and the physics of photoionization in the ISM, S{\sc ii} is the dominant ion. This work demonstrates that there is enough uncertainty in the positions of the resonant S{\sc i} and S{\sc ii} absorption lines to warrant shifting the \citet{Gatuzz2024} cross-sections.
The S{\sc ii} absorption template 
provides a relatively good fit to the interstellar absorption after shifting the energy scale on the template by $+7$ to $8$~eV. This shift is comparable to the inconsistencies among various atomic databases and techniques for predicting atomic energy levels. 

The fit with the S{\sc ii} template alone also exhibits residuals near $2.485$~keV, next to the absorption features from K$\gamma$ and above (Figure~\ref{fig:spectra}). Given the uncertainty in the position of the K$\beta$ feature, it is possible that the positions of the other lines also  require shifting. To explore this possibility, we applied a $-1$~eV shift to the S{\sc ii} template for energies above 2.478~keV, which changes the spacing between the K$\beta$ and higher order lines. This improved the S{\sc ii} fit by $\Delta C = 20$ without affecting other best fit parameters. Shifting the higher order absorption features did not significantly improve the fit statistic when using both the gas and dust templates ($\Delta C = 2$), and the other best fit parameters were within $1\sigma$ of those reported in Table~\ref{tab:dustfit}. 
This leads us to conclude that, while the S{\sc ii} absorption template warrants improvements, it does not affect the final conclusions of this work.

Up until now, S depletion studies of the diffuse ISM have relied solely on comparing gas-phase S abundance with an assumed abundance table, which is tricky because S{\sc ii} lines in the UV are typically saturated and assessing the true abundance relative to H is difficult due to the prevalence of S{\sc ii} in ionized regions of the ISM \citep{Jenkins2009}. Nonetheless, estimates for S depletion can be as high as 30\% in $\log N_H > 19.5$ sight lines \citep{Jenkins2009}. \citet{Psaradaki2024} also used UV absorption from S{\sc ii} to estimate a depletion of 40\% for the $N_H \approx 2 \times 10^{21}$\col\ sight line of LMXB Cygnus~X-2. X-ray absorption around 2.5~keV allows us to capture both atomic and solid forms of S in the ISM. After fitting the XRISM spectrum of \NShigh\ with an atomic S{\sc ii} absorption template, we find residual absorption from interstellar dust. Three Fe-S compounds -- troilite (FeS), pyrrhotite (Fe$_7$S$_8$), and pyrite (FeS$_2$) -- provide relatively equal fit statistic improvements for the S $K$-shell photoabsorption region. The abundance of S solids inferred from each compound also agree. We find that a combined gas and dust fit is able to account for the total abundance of S in the sight line, as inferred from the $K$-shell continuum absorption. 
This work provides the first complete detection of both solid and gas phase S in one sight line, providing a direct measurement of S depletion, which is $40\% \pm 15\%$.

Among various abundance tables, Fe is typically $0.2-0.4$~dex ($1.6 - 2.5$ times) more abundant than S. Depletion fractions for Fe are also typically on the order of 90-99\% \citep{Jenkins2009}. 
The compounds we test all have an Fe:S ratio $\leq 1$, implying that there must be a significant reservoir of solid-phase Fe that is not a sulfide or sulfate. However, studies of Fe $L$-shell photoabsorption features have not identified a significant contribution from Fe-S compounds \citep{Westphal2019, Psaradaki2023, Corrales2024}. At the present time we are not able to test for the presence of S-compounds without Fe in them, but new laboratory measurements are forthcoming. The current measurement provides an upper limit of 25\% for the amount of interstellar Fe bound in Fe-S compounds in the sightline of \NShigh.

Comparing the S abundance obtained from the photoelectric absorption continuum to the S{\sc ii} abundance obtained from the K$\beta$ line could also provide a means to measure depletion, even when the spectral signatures of the dust are not present at high signal-to-noise. 
When we do this for the \NSmodsrc\ sight line, we get a depletion fraction that is consistent with that of \NShigh, but with very large error bars. Stacking the current data with future XRISM measurements, or longer exposures of other obscured X-ray binaries, will yield better results.

\NShigh\ and \NSmodsrc\ are distant X-ray binaries, making them unique targets for studying the integrated properties of the Milky Way ISM. The distance to \NShigh\ is estimated to be $11 \pm 3$~kpc, and the distance to \NSmodsrc\ is constrained by measurements from dust scattering echos to be $11.5 \pm 0.3$~kpc. Both sources are located within 1~degree of the Galactic plane and 20-25~degrees away from the Galactic Center, putting them on the distant side of the Galactic disk. 
The positions of these sources suggest that our current measurements could represent a reasonable average value of S depletion when integrating across all phases of the Milky Way ISM.

\begin{ack}

We thank E. Gatuzz, F. Hasoglu, and T. Gorczycka for useful discussion regarding the S $K$-shell photabsorption features and for making the published atomic S templates publicly available. We also thank E. Behar, R. Tomaru, and C. Done for helpful discussion regarding potential contaminating line features from photoionized plasmas. We also thank the manuscript reviewers, M. Leutenegger and F. Paerels, for insightful comments and questions that greatly enhanced the clarity and quality of the paper.

\end{ack}

\section*{Funding}

This research was supported by NASA grants 80NSSC18K0978, 80NSSC20K0883, and 80NSSC25K7064. Partial support was also provided by NSF award 2205918.
SZ acknowledges support from the Research Fellowship Program of the European Space Agency (ESA). 
RB acknowledges support from NASA grant number 80GSFC24M0006. 
IP is supported by NASA through the Smithsonian Astrophysical Observatory (SAO) contract SV3-73016 to MIT for Support of the Chandra X-Ray Center (CXC) and Science Instruments. CXC is operated by SAO for and on behalf of NASA under contract NAS8-03060.
MS acknowledges support by Grants-in-Aid for Scientific Research 19K14762 and 23K03459 from the Ministry of Education, Culture, Sports, Science, and Technology (MEXT) of Japan. MM acknowledges support from Yamada Science Foundation as well as  JSPS KAKENHI Grant Number JP21K13958. RB acknowledges support by NASA under award number 80GSFC21M0002. RML acknowledges support by NASA under award number 80NSSC23K0635. 
TN acknowledges the support by JSPS KAKENHI Grant Numbers 23H05441 and 23K17695.

\section*{Data availability} 

The data underlying this article are subject to an embargo of 12 months from the completion of the XRISM Performance Verification phase data collection. Once the embargo expires, the data will be available from NASA and JAXA online data archives.

\appendix 
\section*{Einstein A coefficients for SII K$\beta$}

Table~\ref{tab:AtomicData} gives the calculated Einstein~$A$ coefficients from \citet{Palmeri2008} and from the Flexible Atomic Code (FAC) for strong transitions from the ground state of S{\sc ii}. For this work, we use the average of the two values and assume a $10\%$ uncertainty for error propagation. The three resonance transitions making up the K$\beta$ feature are within 0.05~eV of eachother, so we treated them as one line with the Einstein $A$ coefficients summed. All other ground-state transitions have $A < 10^9$~s$^{-1}$, and so were not included in the analysis.

\begin{table}[]
    \tbl{SII K$\beta$ resonant transitions to ground ($^4$S$_{3/2}$)}{
    \begin{tabular}{l c c c}
        \hline\noalign{\vskip3pt}
        Upper Level & \multicolumn{3}{c}{$A_{21}$ ($10^{12}$~s$^{-1}$)} \\ 
            & \citet{Palmeri2008} & FAC & This work \\
        \hline\noalign{\vskip3pt}
        $^4$P$_{5/2}$ & 1.90 & 1.636 & 1.77 \\
        $^4$P$_{3/2}$ & 1.90 & 1.637 & 1.77 \\
        $^4$P$_{1/2}$ & 1.90 & 1.636 & 1.77 \\
        \hline\noalign{\vskip3pt}
    \end{tabular}}
    \label{tab:AtomicData}
\end{table}

\bibliography{moutard_references, export-bibtex, other_references}{}
\bibliographystyle{aasjournal}

\end{document}